\documentclass[longauth]{aa} 

\usepackage{graphicx}
\usepackage{txfonts}
\usepackage{fixltx2e}
\usepackage{color}
\usepackage{lineno}

\newcommand*\patchAmsMathEnvironmentForLineno[1]{%
  \expandafter\let\csname old#1\expandafter\endcsname\csname #1\endcsname
  \expandafter\let\csname oldend#1\expandafter\endcsname\csname end#1\endcsname
  \renewenvironment{#1}%
     {\linenomath\csname old#1\endcsname}%
     {\csname oldend#1\endcsname\endlinenomath}}%
\newcommand*\patchBothAmsMathEnvironmentsForLineno[1]{%
  \patchAmsMathEnvironmentForLineno{#1}%
  \patchAmsMathEnvironmentForLineno{#1*}}%
\AtBeginDocument{%
\patchBothAmsMathEnvironmentsForLineno{equation}%
\patchBothAmsMathEnvironmentsForLineno{align}%
\patchBothAmsMathEnvironmentsForLineno{flalign}%
\patchBothAmsMathEnvironmentsForLineno{alignat}%
\patchBothAmsMathEnvironmentsForLineno{gather}%
\patchBothAmsMathEnvironmentsForLineno{multline}%
}

\usepackage{hyperref}
\usepackage{breakurl}
\def\src{PKS\,1510--089}
\def\fermi{\textit{Fermi}-LAT}
\def\XRT{\textit{Swift}-XRT}
\newcommand{\kom}[1]{#1}

\begin{document}

   \title{Detection of persistent VHE gamma-ray emission from \src\ by the MAGIC telescopes during low states between 2012 and 2017.}
   \titlerunning{Detection of VHE gamma rays from \src\ during low state.}
   \author{
MAGIC Collaboration:
V.~A.~Acciari\inst{1} \and
S.~Ansoldi\inst{2,21} \and
L.~A.~Antonelli\inst{3} \and
A.~Arbet Engels\inst{4} \and
C.~Arcaro\inst{5} \and
D.~Baack\inst{6} \and
A.~Babi\'c\inst{7} \and
B.~Banerjee\inst{8} \and
P.~Bangale\inst{9} \and
U.~Barres de Almeida\inst{9,10}
% ...
}

   \institute {
Inst. de Astrof\'isica de Canarias, E-38200 La Laguna, and Universidad de La Laguna, Dpto. Astrof\'isica, E-38206 La Laguna, Tenerife, Spain
\and Universit\`a di Udine, and INFN Trieste, I-33100 Udine, Italy
\and National Institute for Astrophysics (INAF), I-00136 Rome, Italy
\and ETH Zurich, CH-8093 Zurich, Switzerland
\and Universit\`a di Padova and INFN, I-35131 Padova, Italy
\and Technische Universit\"at Dortmund, D-44221 Dortmund, Germany
\and Croatian MAGIC Consortium: University of Rijeka, 51000 Rijeka, University of Split - FESB, 21000 Split, University of Zagreb - FER, 10000 Zagreb, University of Osijek, 31000 Osijek and Rudjer Boskovic Institute, 10000 Zagreb, Croatia.
\and Saha Institute of Nuclear Physics, HBNI, 1/AF Bidhannagar, Salt Lake, Sector-1, Kolkata 700064, India
\and Max-Planck-Institut f\"ur Physik, D-80805 M\"unchen, Germany
\and now at Centro Brasileiro de Pesquisas F\'isicas (CBPF), 22290-180 URCA, Rio de Janeiro (RJ), Brasil
\and Unidad de Part\'iculas y Cosmolog\'ia (UPARCOS), Universidad Complutense, E-28040 Madrid, Spain
\and University of \L\'od\'z, Department of Astrophysics, PL-90236 \L\'od\'z, Poland
\and Deutsches Elektronen-Synchrotron (DESY), D-15738 Zeuthen, Germany
\and Institut de F\'isica d'Altes Energies (IFAE), The Barcelona Institute of Science and Technology (BIST), E-08193 Bellaterra (Barcelona), Spain
\and Universit\`a di Siena and INFN Pisa, I-53100 Siena, Italy
\and Universit\`a di Pisa, and INFN Pisa, I-56126 Pisa, Italy
\and Universit\"at W\"urzburg, D-97074 W\"urzburg, Germany
\and  Finnish MAGIC Consortium: Tuorla Observatory \kom{(Department of Physics and Astronomy)} and Finnish Centre of Astronomy with ESO (FINCA), University of Turku, \kom{FIN-20014 Turku}, Finland; Astronomy Division, University of Oulu, FIN-90014 Oulu, Finland 
\and Departament de F\'isica, and CERES-IEEC, Universitat Aut\'onoma de Barcelona, E-08193 Bellaterra, Spain
\and Universitat de Barcelona, ICCUB, IEEC-UB, E-08028 Barcelona, Spain
\and Japanese MAGIC Consortium: ICRR, The University of Tokyo, 277-8582 Chiba, Japan; Department of Physics, Kyoto University, 606-8502 Kyoto, Japan; Tokai University, 259-1292 Kanagawa, Japan; RIKEN, 351-0198 Saitama, Japan
% ...
}
\date{Received ; accepted }

 
  \abstract
   {\src\ is a flat spectrum radio quasar strongly variable in the optical and GeV range. 
So far, very-high-energy (VHE\kom{, $>100$\,GeV}) emission has been observed from this source during either long high states of optical and GeV activity or during short flares. }
   {
We search for low-state VHE gamma-ray emission from \src . 
We aim to characterize and model the source in a broad-band context, which would provide a baseline over which high states and flares could be better understood.
}
   {
\src\ has been monitored by the MAGIC telescopes since 2012. 
We use daily binned \fermi\ flux measurements of \src\ to characterize the GeV emission and select the observation periods of MAGIC during low state of activity. 
For the selected times we compute the average radio, IR, optical, UV, X-ray and gamma-ray emission to construct a low-state spectral energy distribution of the source.
The broadband emission is modelled within an External Compton scenario with a stationary emission region through which plasma and magnetic field are flowing.
\kom{We perform also the emission-model-independent calculations of the maximum absorption in the broad line region (BLR) using two different models.}
}
   {The MAGIC telescopes collected 75\,hrs of data during times when the \fermi\ flux measured above 1\,GeV was below $3\times10^{-8}\mathrm{cm^{-2}s^{-1}}$, which is the threshold adopted for the definition of a low gamma-ray activity state. 
The data show a strongly significant ($9.5\sigma$) VHE gamma-ray emission at the level of $(4.27\pm0.61_{\rm stat})\times 10^{-12}\,\mathrm{cm^{-2}s^{-1}}$ above 150\,GeV, a factor 80 smaller than the highest flare observed so far from this object. 
Despite the lower flux, the spectral shape is consistent with earlier detections in the VHE band.
The broad-band emission is compatible with the External Compton scenario assuming a large emission region located beyond the broad line region.
\kom{
  For the first time the gamma-ray data allow us to place a limit on the location of the emission region during a low gamma-ray state of a FSRQ. 
  For the used model of the BLR, the 95\% C.L. on the location of the emission region allows us to place it at the distance $>74\%$ of the outer radius of the BLR.}
}
   {}


   \maketitle
%
\section{Introduction}

\src\ is a bright flat spectrum radio quasar (FSRQ) located at a redshift of $z=0.36$ \citep{ta96}.
It was the second FSRQ to be detected in the very-high-energy (VHE, $>100$\,GeV) range \citep{ab13}. 
The source is monitored by various instruments spanning the full range from radio up to VHE gamma rays \citep[see e.g.][]{ma10, al14, ah17}. 
Similarly to other FSRQs, the GeV gamma-ray emission of \src\ is strongly variable \citep{ab10, br13, sa13, pr17}. 
Multiple optical flares have been observed from \src\ \citep{ll75,za16}\footnote{\kom{see also \protect\burl{http://users.utu.fi/kani/1m/PKS\_1510-089\_jy.html}}}.

Significant VHE gamma-ray emission from \src\ has been observed on a few occasions: during enhanced optical and GeV states in 2009 \citep{ab13} and 2012 \citep{al14} and during short flares in 2015 \citep{ah17, za16} and 2016 \citep{za17}. 
Interestingly, no variability in VHE gamma rays has been observed during  (or between) the high optical/GeV states in 2009 and 2012 \citep{ab13,al14}. 

The GeV state of \src\ can be studied using the \fermi\ all-sky monitoring data. 
MAGIC is a system of two Imaging Atmospheric Cherenkov Telescopes designed for observations of gamma rays with energies from a few tens of GeV up to a few tens of TeV \citep{al16a}.
Since the detection of VHE gamma-ray emission from \src\ in 2012, a monitoring program is being performed with the MAGIC telescopes.
The aimed cadence of monitoring is 2-6 pointings per month, with individual exposures of 1-3\,hrs.
The source is visible by MAGIC 5\,months of the year. 
We use the \fermi\ data to select periods of low gamma-ray emission of \src .
Then, we select a subsample of the MAGIC telescope data taken between 2012 and 2017, and contemporaneous multiwavelength data from a number of other instruments, in order to study the quiescent VHE gamma-ray state of the source.
Such low emission can then be used as a baseline for modeling of flaring states.

In Section~\ref{sec:data} we briefly introduce the instruments that provided multiwavelength data, describe the data reduction procedures and explain the principle of low-state data selection.
In Section~\ref{sec:results} we present the results of the observations, and the broadband emission modelling is illustrated in Section~\ref{sec:model}. 
The most important findings are summarized in Section~\ref{sec:conc}.

\section{Data}\label{sec:data}
\begin{figure*}[h!]
\centering
\includegraphics[width=0.95\textwidth]{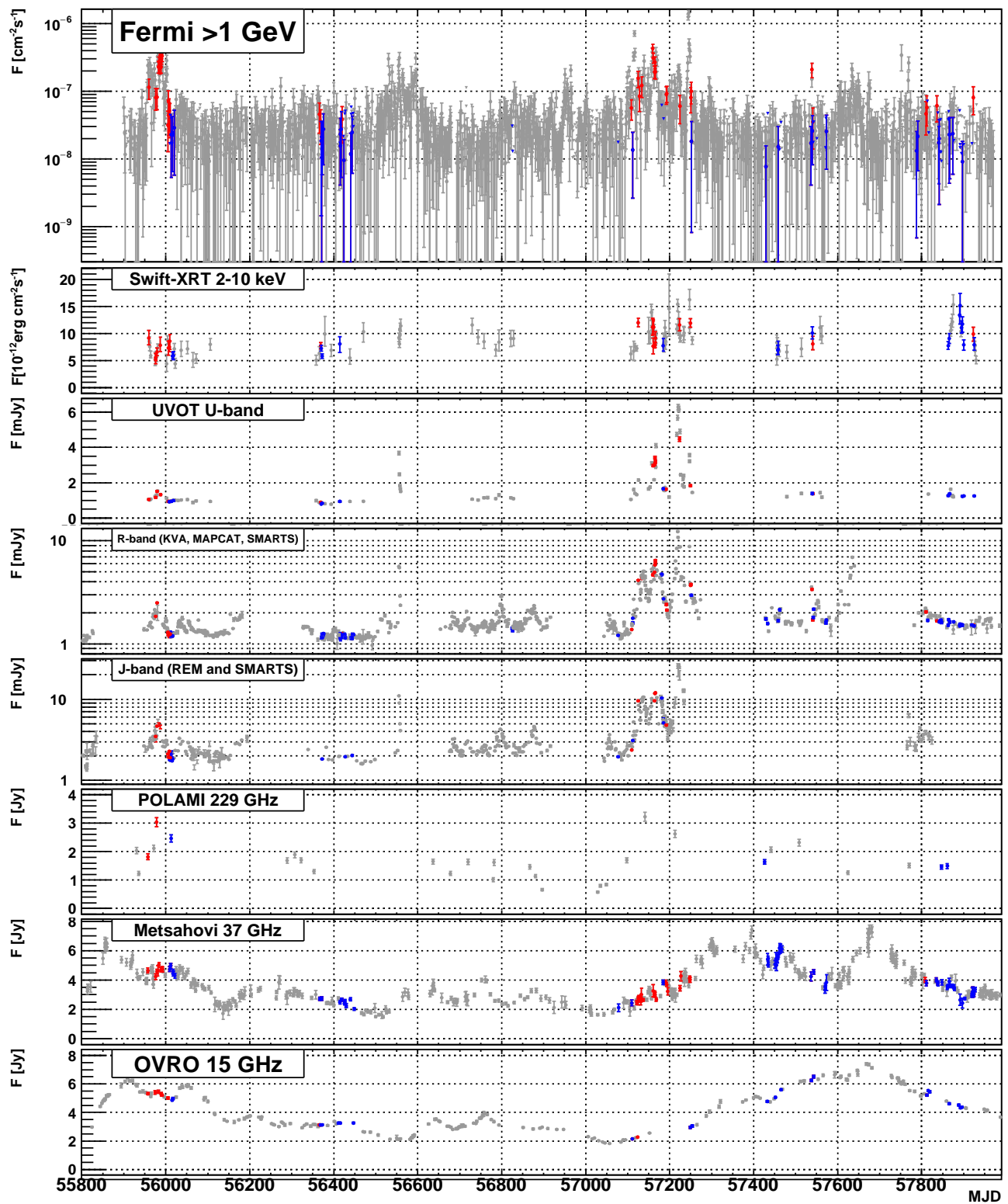}
\caption{
  Multiwavelength light curve of \src\ between 2012 and 2017. From top to bottom:
  \fermi\ flux above 1\,GeV; 
  \XRT\ flux 2-10\,keV; 
  U band flux from UVOT; 
  KVA, SMARTS and MAPCAT optical flux in R-band; 
  IR flux from REM and SMARTS in J band; 
  radio 229\,GHz flux measured by POLAMI; 
  radio 37\,GHz flux measured by Mets\"ahovi; 
  15\,GHz flux measured by OVRO.
The red points show the observations within 12\,h (or 3 days for the radio measurements) when MAGIC data have been taken during the time that \fermi\ flux is above $3\times10^{-8}\mathrm{cm^{-2}s^{-1}}$, while the \kom{blue} points are observations in time bins with \fermi\ flux below this flux value (i.e. the low-state sample).
IR, optical and UV data have been dereddened using \citet{sf11}. 
}\label{fig:lc}
\end{figure*}

The continuous monitoring of \src\ in the GeV band provided by \fermi\ allows us to identify the low emission states of the source.
Multiwavelength light curves from the radio band up to the GeV band are shown in Fig.~\ref{fig:lc}.

\subsection{\fermi}\label{sec:fermi}
\fermi\ monitors the high energy gamma-ray sky in the energy range from 20 MeV to beyond 300 GeV \citep{Atwood09}.
For this work, we have analyzed the Pass 8 SOURCE class events within a region of interest (ROI) of $10^\circ$ radius centered at the position of \src\ in the energy range from 100 MeV to 300 GeV.
A zenith angle cut of $<90^\circ$ was applied to reduce the contamination from the Earth’s limb.
The analysis was performed with the \texttt{ScienceTools} software package version \texttt{v11r7p0} using the \texttt{P8R2\_SOURCE\_V6}\footnote{\burl{http://fermi.gsfc.nasa.gov/ssc/data/analysis/documentation/Cicerone/Cicerone_LAT_IRFs/IRF_overview.html}} instrument response function and the \texttt{gll\_iem\_v06} and \texttt{iso\_P8R2\_SOURCE\_V6\_v06} models\footnote{\burl{http://fermi.gsfc.nasa.gov/ssc/data/access/lat/BackgroundModels.html}} for the Galactic and isotropic diffuse emission \citep{ac16}, respectively.

A first unbinned likelihood fit was performed for the events collected within five months from 01 February to 30 June 2013 (MJD 56324--56474) using \texttt{gtlike}, and including in the model file all 3FGL sources \citep{Acero15} of $20^\circ$ from \src .
We repeat the same 5-month analysis using the preliminary 8-year Source List~\footnote{\burl{https://fermi.gsfc.nasa.gov/ssc/data/access/lat/fl8y/}} instead of the 3FGL catalog to search for bright sources within 20$^\circ$ of \src\ . 
No new strong sources were found. 
The model generated from the 3FGL catalog was used for the subsequent analysis.
As we are interested in short time scale (daily) fluxes of \src\ the purpose of this first fit is to identify weak nearby sources that can be removed from the source model, simplifying it. 
Hence, the sources with a test statistic (TS; \citealp{ma96}) below 5 were removed from the model file.
Next, the optimized output model file was used to produce the \src\ light curve with 1-day time bins above 1 GeV in the full time period from 5 December 2011 to 7 August 2017 (MJD 55900--57972). 
The same optimized output model is later also used for the spectral analysis. 
In the light curve calculations the spectra of \src\ were modeled as power law leaving both the flux normalization and the spectral index as free parameters.
The normalization of the Galactic and isotropic diffuse emission models was left to vary freely during the calculation of both the light curves and the spectrum.
In addition, the spectra of all sources except \src\ and the highly variable source 3FGL 1532.7-1319 (located 6.45$^\circ$ from \src , and having variability index of 1924.7 from the 3FGL catalog) were fixed to the catalog values.

\begin{figure}[t]
\centering
\includegraphics[width=0.49\textwidth]{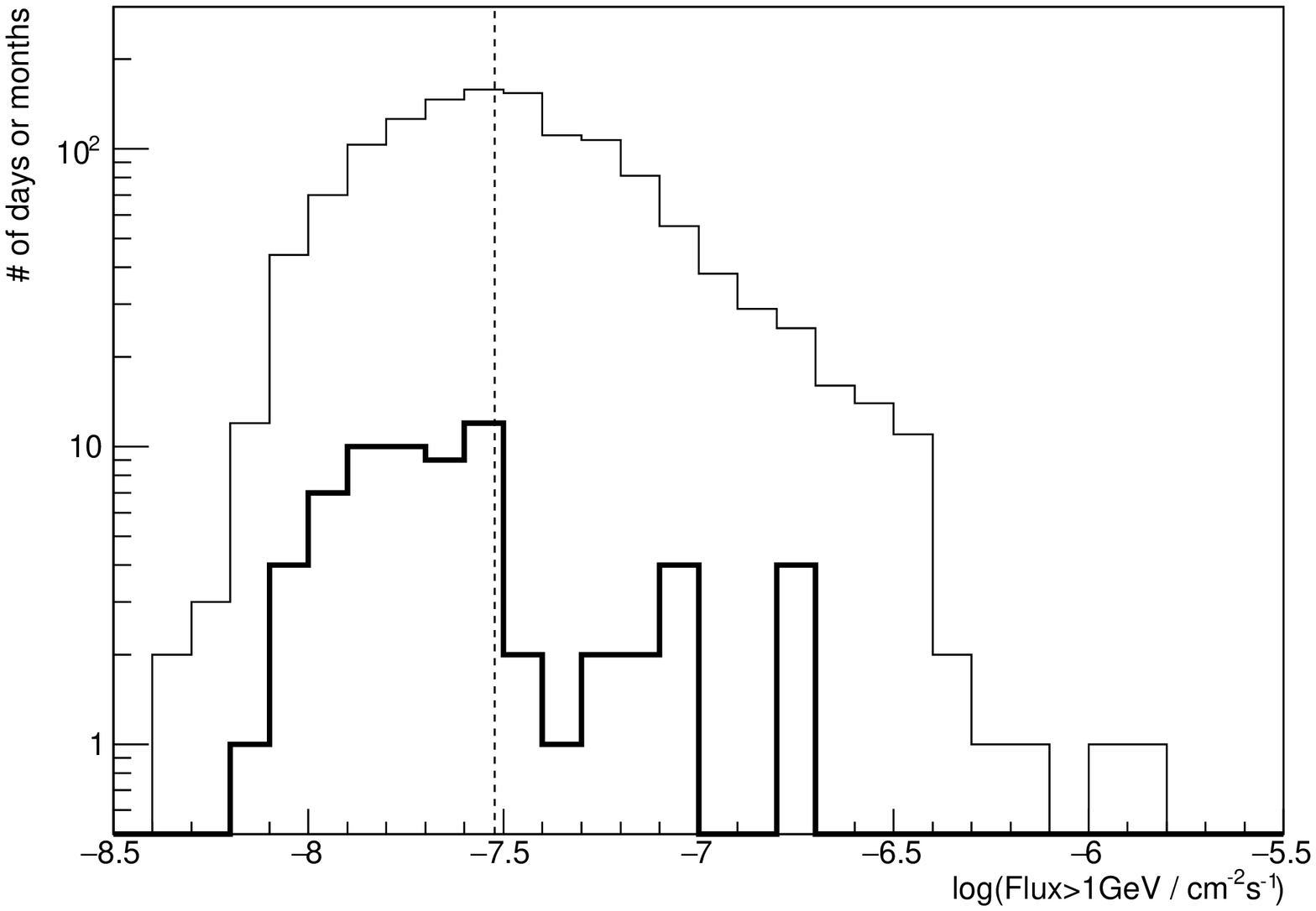}
\caption{
Distribution of flux above 1\,GeV measured with \fermi\ in 30-day (thick line) or 1-day (thin line) bins. 
The vertical dashed line shows the value of the cut separating the low state. 
}\label{fig:fermidist}
\end{figure}

In order to estimate when the flux can be considered being in a low state, we first calculate a light curve in relatively wide bins of 30 days in the full time period. 
This allows us to estimate the flux with relative uncertainty $\lesssim20\%$ for all the points and hence disentangle intrinsic variability from the fluctuations of the measured flux induced by statistical uncertainties. 
In Fig.~\ref{fig:fermidist} we present the distribution of the flux above 1\,GeV, which shows that during the low state the flux is between (1--3) $\times10^{-8}\,\mathrm{cm^{-2}s^{-1}}$ in contrast to the value of  $>3\times 10^{-8} \mathrm{cm^{-2}s^{-1}}$ during active (flaring) periods.
Hence, we select the days of low state if: 
\begin{equation}
F_{>1\mathrm{GeV}}< 3\times 10^{-8} \mathrm{cm^{-2}s^{-1}}. \label{eq:lowstate}
\end{equation}
The cut separates the low \kom{flux} peak of the daily fluxes distribution from the power-law extension of the high-flux days (see Fig.~\ref{fig:fermidist}). 
The effect that choosing a different energy threshold would have on the data selection is discussed in Appendix~\ref{app:datasel}. 
We note, however, that due to the low state and short exposure times the flux measurements during single nights are quite uncertain. 
The typical uncertainty of the flux in those time bins is $\sim1.5\times 10^{-8} \mathrm{cm^{-2}s^{-1}}$. 
We also include in the low-state sample nights for which the \fermi\ flux did not reach TS of 4.
The average 95\% C.L. flux upper limit on those nights is also $\sim3\times 10^{-8} \mathrm{cm^{-2}s^{-1}}$.

For the low-state spectrum we combine individual one day integration windows selected with flux fulfilling Eq.~\ref{eq:lowstate}. 
The spectrum, calculated above 100 MeV, is well described as a power law with spectral index $2.56\pm0.04$, with a TS$=1656.0$. 
The possible curvature in the spectrum is investigated by fitting the spectrum with a logparabola which yields a TS$=1655.42$ and negligible curvature ($\beta=0.06\pm0.04$). 
Therefore, no hint of spectral curvature was found during the low-state periods considered in this analysis. 
The selection of \fermi\ observation days according to the flux $>1$\,GeV can bias the reconstructed average spectrum in this energy range. 
To investigate such possible bias for the selected low-emission time periods we also calculate the spectral index in the energy range 0.1--1\,GeV \kom{(not affected by the data selection)} and obtain $2.41\pm0.06$. 
Moreover, the \fermi\ spectral energy points above 1\,GeV are $\sim 25\%$ lower than the extrapolation of the spectrum below 1\,GeV\kom{, suggesting that indeed there is an up to $25\%$ underestimation bias in the obtained \fermi\ flux above 1\,GeV}.

\subsection{MAGIC}
MAGIC is a system of two imaging atmospheric Cherenkov telescopes.
The telescopes are located in the Canary Islands, on La Palma ($28.7^\circ$\,N, $17.9^\circ$\,W), at a height of 2200 m above sea level \citep{al16a}.
The large mirror dish diameter of 17\,m, resulting in low energy threshold, makes it a perfect instrument for studies of soft-spectrum sources such as FSRQs. 
As \src\ is a southern source, only observable at zenith angle $>38^\circ$, the corresponding trigger threshold \kom{would be} $\gtrsim 90$\,GeV for a Crab nebula-like spectrum \citep{al16b}, about 1.7 times larger than for the low zenith observations.
\kom{
About 70\% of the data of \src\ was taken at the culmination, with zenith angle $<40^\circ$.
Moreover, \src\ is intrinsingly soft; the analysis energy threshold is only $\sim 80$\,GeV for a source with a spectral index of $\sim -3.3$.
Note also that the energy threshold of Cherenkov telescopes is not a sharp one and the unfolding procedure allows us to reconstruct the source spectrum slightly below the nominal value of the threshold.} 

\begin{figure}[t]
\centering
\includegraphics[width=0.49\textwidth]{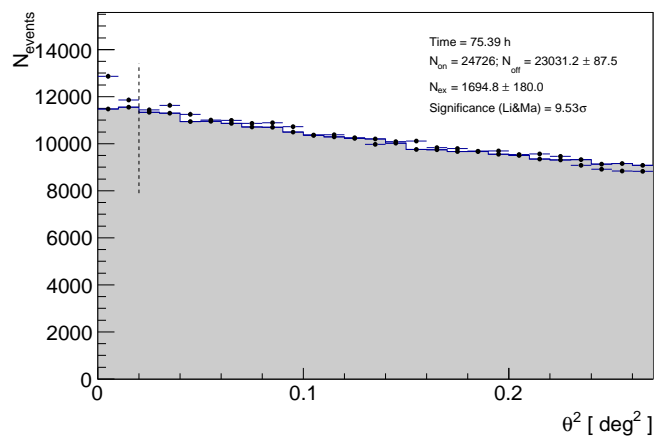}
\caption{
  Distribution of $\theta^2$, the squared angular distance between the reconstructed arrival direction of individual events and the nominal source position (points) or background estimation region (gray filled area) for MAGIC observations of \src .
  Dashed line shows the value of the $\theta^2$ up to which the significance of the detection (see the inset text) is calculated. 
}\label{fig:th2}
\end{figure}

Between 2012 and 2017 the MAGIC telescopes observed \src\ during 151 nights, out of which 115 passed at least partially the data quality selection cuts. 
We then select the nights corresponding to the \fermi\ periods fulfilling the Eq.~\ref{eq:lowstate} condition.
Such procedure results in low-state data stacked from 76 nights, amounting to a total observation time of 75\,hrs. 
The cut on the flux $>1$\,GeV excludes the MAGIC data reporting the detections of the two flares observed in 2015 \citep{ah17} and 2016 \citep{za17}, as well as most of the data used for the detection during the high state of 2012 \citep{al14}. 
The data were analyzed using MARS, the standard analysis package of MAGIC \citep{za13, al16b}.
Due to evolving  telescope performance the data have been divided into 6 analysis periods. 
Within each analysis period proper Monte Carlo simulations are used for the analysis. 
At the last stage the analysis results from all the periods are merged together.
This low-state data set shows a gamma-ray excess with a significance of $9.5\sigma$ (see Fig.~\ref{fig:th2}).

\subsection{\XRT}
Since 2006, the source is monitored in the X-ray band by the XRT instrument on board the \textit{Neil Gehrels Swift Observatory} \citep[\textit{XRT},][]{2004SPIE.5165..201B}.
In total, 243 raw images are publicly available in SWIFTXRLOG (\textit{Swift}-XRT Instrument, Log)\footnote{\burl{https://heasarc.gsfc.nasa.gov/W3Browse/swift/swiftxrlog.html}}. 
From those we selected 17 images based on the simultaneity to the GeV low-flux state and contemporaneousness with MAGIC observations.
\kom{The standard \XRT\ analysis\footnote{\burl{http://www.swift.ac.uk/analysis/xrt/}} is described in detail in \cite{ev09}. 
The data} are processed following the procedure described by \citet{2017A&A...608A..68F}, assuming a fixed equivalent Galactic hydrogen column density $n_H = 6.89 \times 10^{20}\,\rm cm^{-2}$ reported by \citet{2005A&A...440..775K}.
We defined the source region as a circle of 20 pixels ($\sim47$'') centered on the source, and a background region as a ring also centered on the source with inner and outer radii of 40 ($\sim$94'') and 80 pixels ($\sim$188''), respectively.

In order to calculate the average low-state X-ray spectrum of \src\ we have combined all selected individual \kom{\XRT{}} pointings (see the \kom{blue} points in Fig.~\ref{fig:lc}), adding up to a total exposure of 30\,ks.
The 2--10 keV\,flux measured during those 17 pointings shows clear variability.
Fitting the flux with a constant value yields $\chi^2/N_{\rm dof}=84/16$; however, the amplitude of the variability is moderate (the RMS of the points is about $\sim30\%$ of the mean flux).
The average spectrum is well fitted ($\chi^2/N_{\mathrm{dof}} = 187.7/214$) with a power law with an index of $1.382\pm0.020$ and $F_{2-10\,\mathrm{keV}} = 8.14^{+0.25}_{-0.19}\times 10^{-12}\mathrm{\,erg\,cm^{-2}s^{-1}}$.
The spectral index does not show \kom{significant} variability  (fit to constant yields $\chi^2/N_{\mathrm{dof}}$=31.19/16, which translates to a chance probability of $\sim 1.2\%$).
A harder-when-brighter trend is \kom{only} hinted, with a Pearson's rank coefficient for a linear correlation between flux and spectral index of 0.81 (2-10 keV) and 0.74 (0.3-10 keV). 

\subsection{Optical observations}
\src\ is regularly monitored as part of the Tuorla blazar monitoring program\footnote{\url{http://users.utu.fi/kani/1m}} in the R band using a 35\,cm Celestron telescope attached to the KVA (Kunglinga Vetenskapsakademi) telescope located at La Palma. 
The monitoring covers the period of 2012-2017 and the observations were mostly contemporaneous with the MAGIC observations of the source. 
The data analysis was performed with the semi-automatic pipeline using the standard analysis procedures (Nilsson et al. in prep). 
The differential photometry was performed using the comparison star magnitudes from \citet{villata97}.

Calar Alto data were acquired as part of the MAPCAT (Monitoring AGN with Polarimetry at the Calar Alto 2.2m Telescope) project\footnote{\url{http://www.iaa.es/~iagudo/_iagudo/MAPCAT.html}}, see \citet{ag12}.
  The MAPCAT data presented here were reduced following the procedure explained in \citet{jo10}.

Additionally, we  used the publicly available data in the R band from the Small and Moderate Aperture Research Telescope System (SMARTS) instrument located at Cerro Tololo Interamerican Observatory (CTIO) in Chile \citep{bo12}, processed as described in \citet{ah17}.
  
The KVA, MAPCAT and SMARTS R-band data have been corrected for Galactic extinction using $\mathrm{A_R} = 0.217$ \citep{sf11}. 
In the optical range, \src\ shows mostly low emission throughout 2012--2014 and during 2017.  
Strong flares are seen in 2015 and 2016 at the times of high GeV state. 
The individual measurements performed in the optical range have very small statistical uncertainties, well below the variability observed during the selected low-state nights. 
Therefore, for the modelling, we use the average optical flux from 53 nights of observations (47 nights of observations with KVA, 3 with MAPCAT and 13 with SMARTS).
We take as the uncertainty the RMS spread of the measurements. 
By applying such a procedure, we obtain that the mean optical flux during the low state is $1.55\pm 0.57$\,mJy.  
\kom{In band B we combine \textit{Swift}-UVOT data (see the next section) with the SMARTS data, bringing the total number of observing nights to 20 and average flux of $1.22\pm 0.46$\,mJy}. 

\subsection{\textit{Swift}-UVOT}
The Ultraviolet/Optical Telescope (UVOT, \citealp{po08}) is an instrument on board the \textit{Swift} satellite operating in the 180--600\,nm wavelength range.
The source counts were extracted from a circular region centered on the source with 5'' radius, the background counts from an annular region centered on the source with inner and outer radius of 15'' and 25'' respectively. 
The data calibration was done following \cite{ra10}, where the effective wavelength, counts-to-flux conversion factor, and Galactic extinction for each filter were calculated in an iterative procedure by taking into account the filter's effective area and the source's spectral shape.
\kom{The Galactic extinction values derived from the re-calibration procedure are 
$A_v=0.28$\,mag, 
$A_b=0.37$\,mag, 
$A_u=0.44$\,mag, 
$A_{w1}=0.63$\,mag, 
$A_{m2}=0.78$\,mag, 
$A_{w2}=0.74$\,mag.} 
The variability of the UV flux during the low-state nights is rather minor. 
The average flux of the quiescent state was derived in the same way as for optical data.
The number of quasi-simultaneous UVOT observations, contemporaneous to MAGIC observations during the \fermi\ low gamma-ray state is 9--13, depending on the filter.

\subsection{IR}
We use infrared observations of \src\ performed with the REM (Rapid Eye Mount, \citealp{ze01, co04}) 60 cm diameter telescope located at La Silla, Chile.
The observations were performed with J, H and Ks filters, with individual exposures ranging from 12\,s to 30\,s. 
After calibration using neighbouring stars, the magnitudes were converted to fluxes using the zero magnitude fluxes from \cite{me90}. 
Additionally, we used the publicly available data in the J and K bands from SMARTS \citep{bo12}, processed as described in \citet{ah17}.

Since the data were taken independently from MAGIC, a limited number of nights of MAGIC observations have quasisimultaneous REM or SMARTS data.
The data taken during the times classified as low state consist of 5 nights of REM data for H filter and 13 nights of REM or SMARTS data from J and K filters.
Moreover, one of the nights observed by SMARTS on MJD 57181 had a major IR flare where the flux increased by a factor of $\sim$5--6 with respect to the average flux value of the rest of the selected data. 
We nevertheless apply the same procedure as for R-band KVA data, averaging the IR flux over those low-state observations, neglecting, however, the night of the IR flare.
We obtain $F_K=7.3\pm 2.7$\,mJy, $F_H=4.2\pm2.4$\,mJy, $F_J=2.3\pm1.0$\,mJy.
Including the night of the IR flare in the sample would change the $F_J$ and $F_K$ fluxes relatively mildly ($\sim 30\%$ increase), however it would increase the RMS considerably to a value comparable to the flux. 

\subsection{Radio}
We use radio monitoring observations of \src\ performed by OVRO (15\,GHz, \citealp{ri11}), Mets\"ahovi (37\,GHz, \citealp{te98}) and POLAMI (86\,GHz, 229\,GHz). 
We also use CARMA data taken at 95\,GHz between August 2012 and November 2014 \citep{ra16}.

POLAMI (Polarimetric Monitoring of AGN at Millimetre Wavelengths)\footnote{\url{http://polami.iaa.es}} \citep{ag18a,ag18b, th18} is a long-term program to monitor the polarimetric properties (Stokes I, Q, U, and V) of a sample of $\sim$40 bright AGN at 3.5 and 1.3 millimeter wavelengths with the IRAM 30m Telescope near Granada, Spain.
The program has been running since October 2006, and it currently has a time sampling of $\sim$2 weeks.
The XPOL polarimetric observing setup has been routinely used as described in \cite{th08} since the start of the program.
The reduction and calibration of the POLAMI data presented here are described in detail in \cite{ag10,ag14,ag18a}.

The 37 GHz observations were made with the 13.7 m diameter Aalto University Mets\"ahovi radio telescope\footnote{\url{http://metsahovi.aalto.fi/en/}}, which is a radome enclosed Cassegrain type antenna situated in Finland. 
The measurements were made with a 1 GHz-band dual beam receiver centered at 36.8 GHz. 
The HEMPT (High Electron Mobility Pseudomorphic Transistor) front end operates at room temperature. 
The observations are Dicke switched ON--ON observations, alternating between the source and the sky in each feed horn. 
A typical integration time to obtain one flux density data point is between 1200 and 1800 s. 
The detection limit of the telescope at 37 GHz is of the order of 0.2\,Jy under optimal conditions. 
Data points with a signal-to-noise ratio $<4$ are considered non-detections. 
The flux density calibration is set by observations of the HII region DR 21. 
The sources NGC 7027, 3C 274 and 3C 84 are used as secondary calibrators. 
A detailed description of the data reduction and analysis is given in \cite{te98}. 
The error estimate in the flux density includes the contribution from the measurement RMS and the uncertainty of the absolute calibration. 

The Owens Valley Radio Observatory (OVRO) 40-Meter Telescope uses off-axis dual-beam optics and a cryogenic receiver with 3~GHz bandwidth centered at 15~GHz. 
Atmospheric and ground contributions as well as gain fluctuations are removed with the double switching technique \citep{1989ApJ...346..566R} where the observations are conducted in an ON-ON fashion so that one of the beams is always pointed on the source. 
Until May 2014 the two beams were rapidly alternated using a Dicke switch. 
Since then a new pseudo-correlation receiver replaced the old one, and a 180$^\circ$ phase switch is used. 
Relative calibration is obtained with a temperature-stable noise diode to compensate for gain drifts. 
The primary flux density calibrator for those observations was 3C~286 with an assumed value of 3.44~Jy\citep{1977A&A....61...99B}, while DR21 is used as a secondary calibrator source. 
Details of the observation and data reduction procedures are given in \citet{ri11}.

The radio flux at all frequencies shows slow variability, not simultaneous with the flares observed at higher energies. 
In order to obtain the average emission during the low gamma-ray state we apply the same procedure as for the R-band flux; however we apply a larger margin in time, using the data within $\pm$3 days from the MAGIC observations during low \fermi\ flux. 
We obtain
$F_{\mathrm{15\,GHz}}=4.4\pm 1.2$\,Jy (average over 22 observations),
$F_{\mathrm{37\,GHz}}=3.9\pm 1.1$\,Jy (59 observations),
$F_{\mathrm{86\,GHz}}=3.14\pm0.86$\,Jy (6 observations),
$F_{\mathrm{95\,GHz}}=2.16\pm0.13$\,Jy (9 observations),
$F_{\mathrm{229\,GHz}}=1.76\pm0.42$\,Jy (4 observations).

\section{Low gamma-ray state of \src}\label{sec:results}

\begin{figure}[t]
\centering
\includegraphics[width=0.49\textwidth]{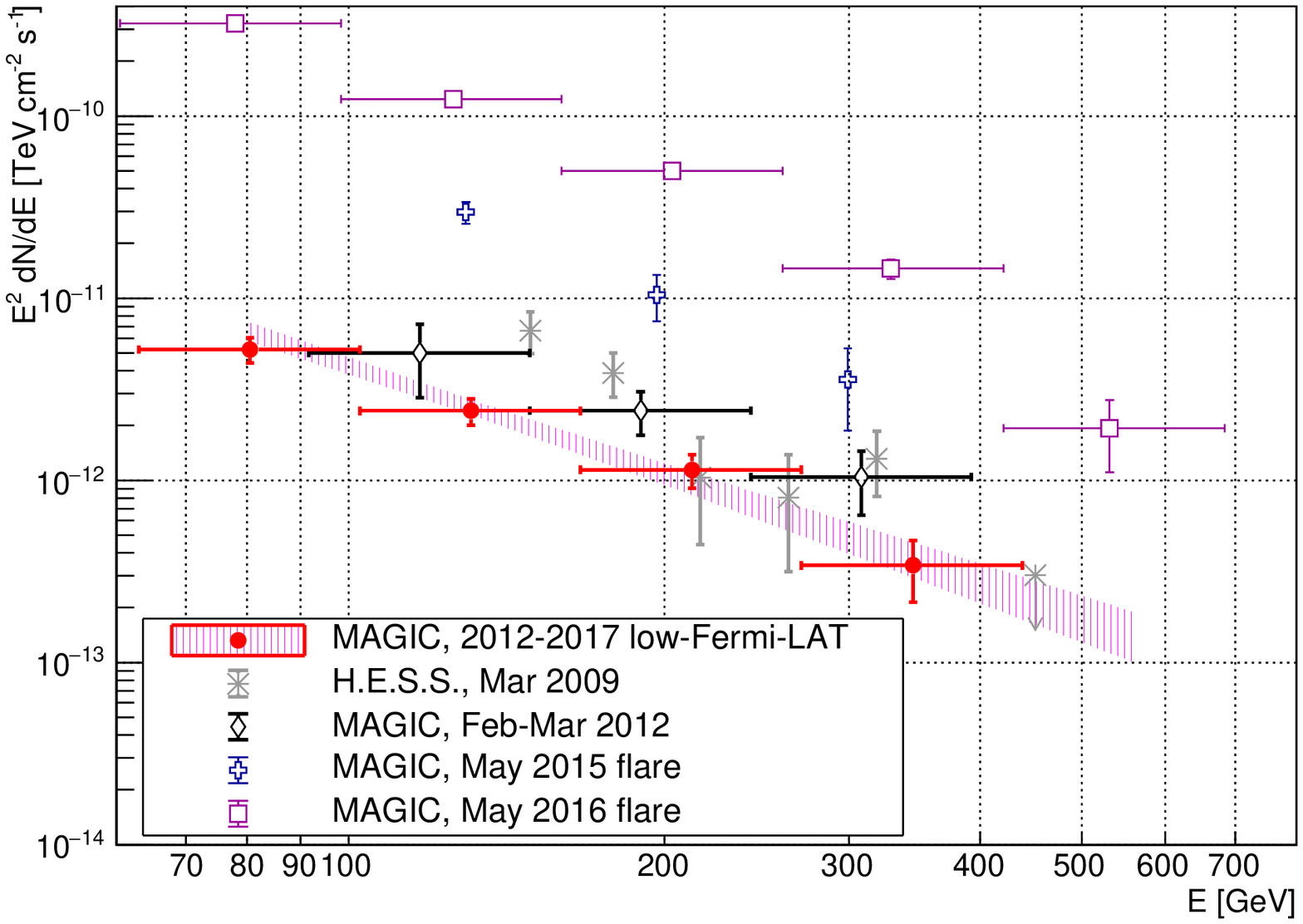}
\caption{
Spectral Energy Distribution (SED) of \src\ during the low state (red filled points and shaded magenta region) compared to historical measurements (open symbols):
high state in March 2009 (gray stars, \citealp{ab13}), high state in February-March 2012 (black diamonds, \citealp{al14}), flare in May 2015 (blue crosses, \citealp{ah17}) and flare in May 2016 (magenta squares, \citep{za17}).
The spectra are not deabsorbed from the EBL extinction. 
}\label{fig:sed}
\end{figure}

The low-state spectrum of \src\ observed by the MAGIC telescopes was reconstructed between 63 and 430\,GeV and is shown in Fig.~\ref{fig:sed}.
The observed spectrum can be described by a power law:
$dN/dE = (4.66\pm0.59_{\mathrm{\kom{stat}}})\times 10^{-11}(E/175\,\mathrm{GeV})^{-3.97\pm0.23_{\mathrm{\kom{stat}}}} \mathrm{cm^{-2}s^{-1}TeV^{-1}}$. 
Correcting for the absorption due to the interaction with the extragalactic background light (EBL) according to \cite{do11}, we obtain the following intrinsic spectrum:
$dN/dE = (7.9\pm1.1_{\mathrm{\kom{stat}}})\times 10^{-11}(E/175\,\mathrm{GeV})^{-3.26\pm0.30_{\mathrm{\kom{stat}}}} \mathrm{cm^{-2}s^{-1}TeV^{-1}}$.
\kom{Since the excess to residual background ratio is of the order of 6\%, the systematic uncertainty on the flux normalization (without the effect of the energy scale) is $\pm$20\%, larger than for typical MAGIC observations, \citep{al16b}.
  Also the systematic uncertainty on the spectral slope is increased by the low excess to residual background ratio and following the prescription of Section~5.1 in \cite{al16b} can be estimated as $\pm0.4$.
The uncertainty of the energy scale is $\pm15\%$. }
Comparing with previous measurements, the high-state detection in 2012 \citep{al14} gives $\sim 1.7$ times larger flux than the low state studied here. 
On the other hand, the most luminous flare observed from \src\ in May 2016 gives flux $\sim$40--80 times higher than the low state (for the MAGIC and H.E.S.S. observation window respectively, see \citealp{za17}). 
Interestingly, the intrinsic spectral index of $-3.26\pm0.30_{\mathrm{stat}}$ is consistent within the uncertainties with the one obtained during the high state in the 2012  ($-2.5\pm0.6_{\mathrm{stat}}$, \citealp{al14}),  the 2015 flare ($-3.17\pm0.80_{\mathrm{stat}}$, \citealp{ah17}) and the 2016 flare ($-2.9\pm0.2_{\mathrm{stat}}$, $-3.37\pm0.09_{\mathrm{stat}}$, \citealp{za17}).

\begin{figure*}[t]
\centering
\includegraphics[width=0.98\textwidth]{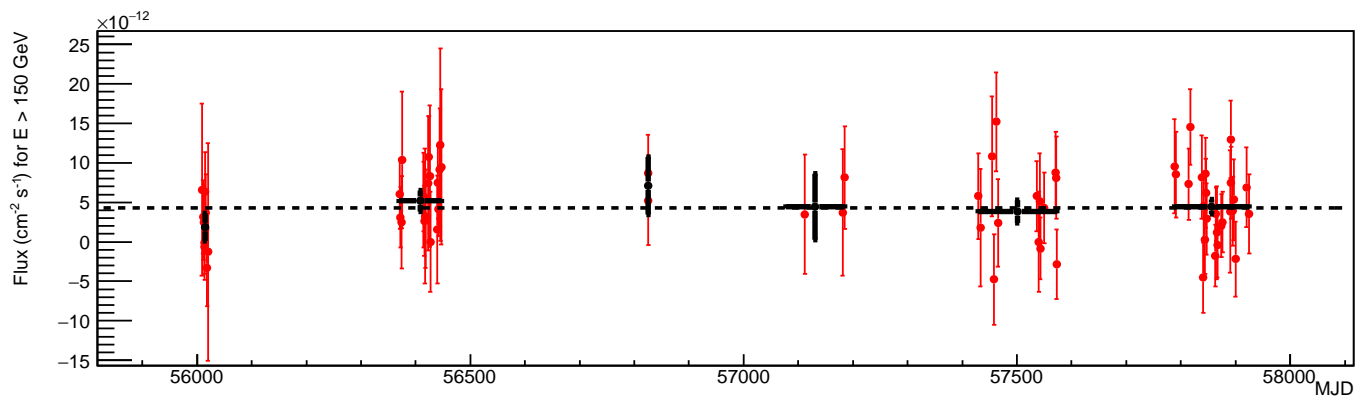}
\caption{
Light curve of \src\ obtained with MAGIC observations during the low state in daily (red thin lines) and yearly (black thick lines) binning.
For clarity three nights with short exposure (resulting in flux estimation uncertainty $>15\times10^{-12}\mathrm{cm^{-2}s^{-1}}$) are omitted from the plot.
}\label{fig:magiclc}
\end{figure*}

As reported in Section~\ref{sec:data}, the IR to UV low-state data show variability at the level of $\sim40$\%. 
We search for possible variability in the MAGIC data taken during the defined low state by computing light curves using different binnings (see Fig.~\ref{fig:magiclc}). 
Both the daily ($\chi^2/\mathrm{N_{dof}} = 51.9/74$) and yearly ($\chi^2/\mathrm{N_{dof}} = 3.08/5$) light curves do not show any evidence of variability when fitted with a constant flux model. 
The gamma-ray flux of \src\ during the low state is, however, too weak for probing variability with a similar relative amplitude at GeV energies with MAGIC or \fermi\ as observed in IR-UV. 
The average emission of the low state above 150\,GeV is $(4.27\pm0.61_{\rm stat})\times 10^{-12}\,\mathrm{cm^{-2}s^{-1}}$, which is also below the all-time average of the H.E.S.S. observations ($(5.5\pm0.4_{\rm stat})\times 10^{-12}\,\mathrm{cm^{-2}s^{-1}}$, \citealp{za16}). 

\section{Modelling}\label{sec:model}
\begin{figure*}[t]
\centering
\includegraphics[width=0.49\textwidth]{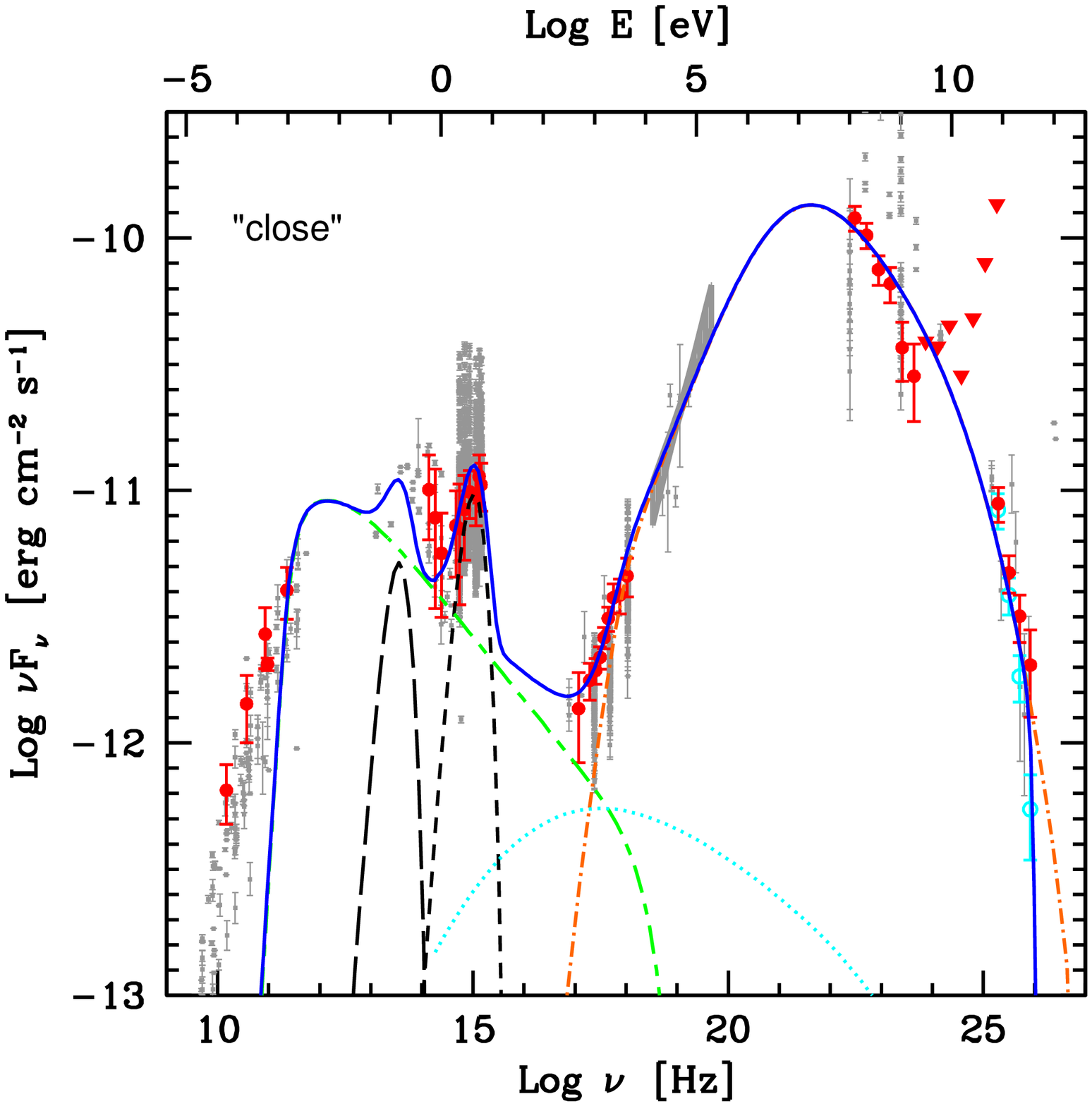}
\includegraphics[width=0.49\textwidth]{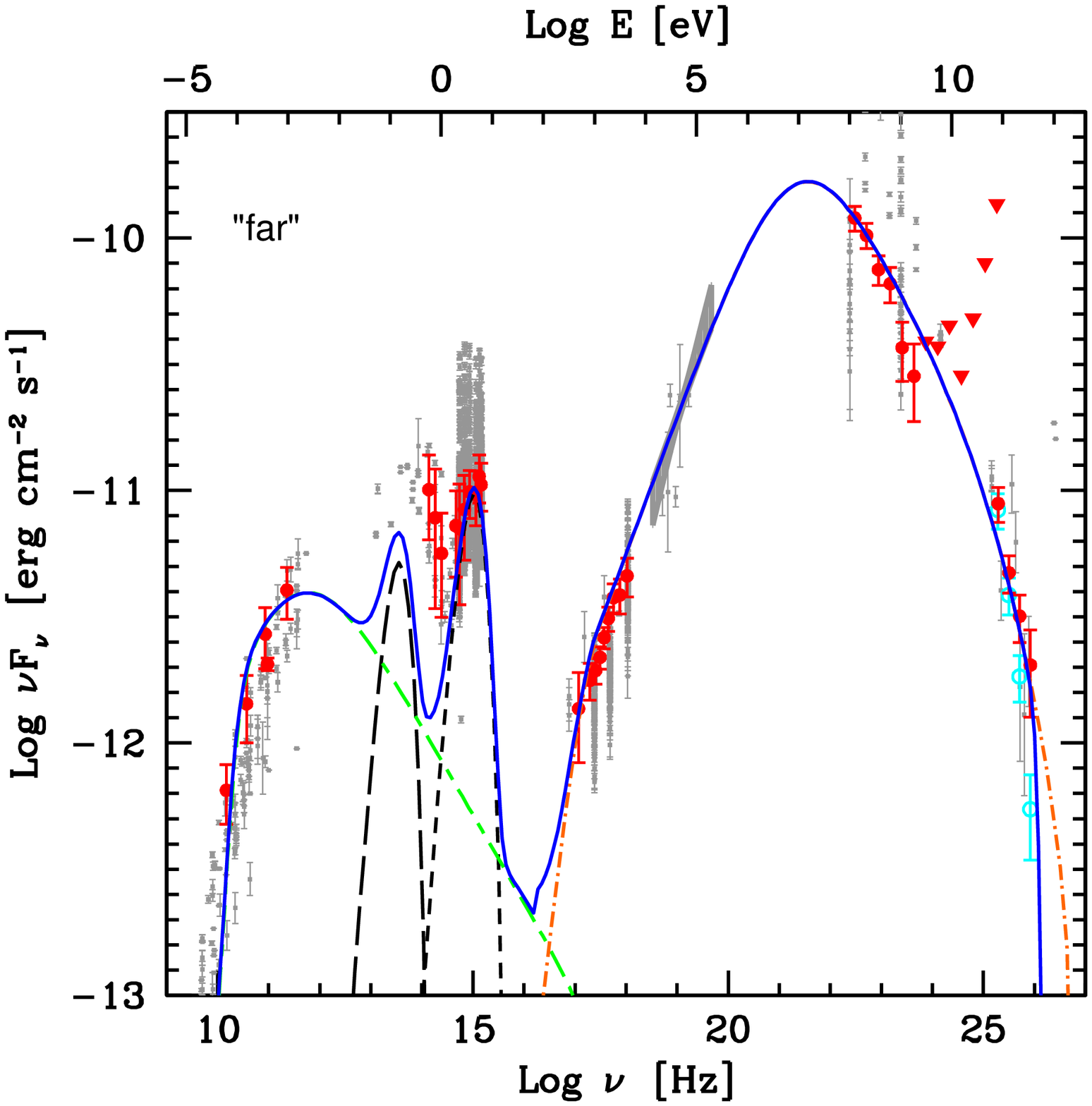}
\caption{
Multiwavelength SED of \src\ obtained from the data contemporaneous to MAGIC observations performed during \fermi\ low state (red points). 
The gray band shows the \textit{Swift}-BAT 105 months average spectrum \citep{oh18}.
Gray dot markers show the historical data from SSDC (\protect\url{www.asdc.asi.it}). 
IR optical and UV data have been dereddened, MAGIC data have been corrected for the absorption by the EBL according to \citet{do11} model.
Observed MAGIC spectral points are shown in cyan. 
The green short-long-dashed curve shows the synchrotron component, and orange dot-dashed the EC component.
SSC component is shown in cyan dotted line.
The long dashed and short dashed black lines show the dust torus and accretion disk emission respectively.
The solid blue line shows the total emission (including absorption in EBL).
The left panel is for the ``close'' model, the right panel for the ``far'' model (see the text).
}\label{fig:mwlsed}
\end{figure*}

The multiwavelength SED constructed from the data selected according to the low flux above 1\,GeV, taken between 2012 and 2017 is shown in Fig.~\ref{fig:mwlsed}. 
We model the broad-band emission using an External Compton scenario (see, e.g., \citealp{sbr94,gh10}) in which the gamma-ray emission is produced due to inverse Compton scattering of a radiation field external to the jet by electrons located in an emission region inside the jet. 
We use a particular scenario applied already to model a high state and a flare from \src\ \citep{al14,ah17}, with the external photon field being the accretion disk radiation reflected by the broad line region (BLR) and dust torus (DT).

We apply the same BLR and DT parameters as in \cite{ah17}, namely a radius of $R_{\rm BLR}=2.6\times 10^{17}$\,cm and $R_{\rm IR}=6.5\times 10^{18}$\,cm respectively. 
BLR and DT reflect $f_{\rm BLR}=0.1$ and $f_{\rm DT}=0.6$ respectively (the so-called covering factor) part of the accretion disk radiation, $L_{\rm disk}=6.7\times 10^{45}$ erg s$^{-1}$.
The DT temperature is set to 1000\,K. 
The emission region is located at the distance $r$ from the base of the jet and has a radius of $R$. 
As in the model employed in \cite{ah17}, jet plasma flows through the emission region. 
The lack of strong variability of the low-state emission and the fact that it reaches sub-TeV energies suggests that the emission region should be beyond the BLR.
At such distances the cross section of the jet is large, making difficult to explain any short-term variability, but the absorption on the BLR radiation is negligible. 
We consider two scenarios for the location of the emission region: the ``close'' scenario with $r=7\times 10^{17}$\,cm and ``far'' scenario with $r=3\times 10^{18}$\,cm. 
In both cases, the dominating radiation field comes from the DT. 
Such distances of the emission region have been applied for modelling of the 2015 flare \citep{ah17} and 2012 high state \citep{al14} respectively.
The size of the emission region $R=2\times 10^{16}$\,cm (for the ``close'' scenario) and $R=3\times 10^{17}$\,cm (for the ``far'' scenario) is on the order of the cross section of the jet at the distance $r$. 
Although it is not a dominant emission component, the model also calculates the synchrotron-self-Compton emission of the source. 

The model parameters for both scenarios are summarized and compared with earlier modelling in Table~\ref{tab:param}.
\begin{table*}
\centering
\begin{tabular}{cccccccccccc}
\hline
\hline
 & $\gamma _{\rm min}$ & $\gamma _{\rm b}$& $\gamma _{\rm max}$& $n_1$& $n_2$ &$B$ &$K$ & $\delta$ & $\Gamma$ & $r$ & $R$ \\
\hline
Low State (close)&$2.5$& $130$         & $3\times 10^{5}$ & $1.9$    & $3.5$ & $0.35$ & $3\times 10^{4}$  & $25$ & $20$ & $7.0\times 10^{17}$ & $2.0\times 10^{16}$ \\
Low State (far)  &$2$  & $300$         & $3\times 10^{5}$ & $1.9$    & $3.7$ & $0.05$ & $80$             & $25$ & $20$ & $3.0\times 10^{18}$ & $3.0\times 10^{17}$ \\
\hline                                                                                                                                                                 
2012           &$3$ & $900$           &$6.5\times 10^{4}$& $1.9$    & $3.85$& $0.12$ & $20$               & $20$ &$20$& $3.1\times 10^{18}$ & $3.0\times 10^{17}$ \\      
2015, Period A &$1$ & $150$ \& $800$  & $4\times 10^{4}$ & 1 \& $2$ & $3.7$ & $0.23$ & $3.0\times 10^{4}$  & $25$ & $20$ & $6.0\times 10^{17}$ & $2.8\times 10^{16}$ \\      
2015, Period B &$1$ & $150$ \& $500$  & $3\times 10^{4}$ & 1 \& $2$ & $3.7$ & $0.34$ & $2.6\times 10^{4}$  & $25$ & $20$ & $6.0\times 10^{17}$ & $2.8\times 10^{16}$ \\      
\hline
\hline
\end{tabular}
\caption{
  Input model parameters for the EC models of \src\ emission for the low state in ``close'' and ``far'' scenario.
  For comparison, model parameters obtained from the 2012 high state \citep{al14} and 2015 flare \cite{ah17} are also quoted.
  The individual columns are  minimum, break and maximum electron Lorentz factor ($\gamma _{\rm min}$, $\gamma _{\rm b}$, $\gamma _{\rm max}$ respectively),
  slope of the electron energy distribution below and above $\gamma_b$ ($n_1$ and $n_2$ respectively), 
  magnetic field in G ($B$),
  normalization of the electron distribution in units of cm$^{-3}$ ($K$),
  Doppler, Lorentz factor, distance and radius of the emission region ($\delta$, $\Gamma$, $r$ (in cm), $R$ (in cm) respectively).
}
\label{tab:param}
\end{table*}
However, the sets of parameters are not unique solutions for describing the low-state SED, as a certain degree of parameter degeneracy occurs in these kind of models (see e.g. synchrotron-self-Compton (SSC) model parameters degeneracy discussed in \citealp{ah17b}).

The data are compared with the model in Fig.~\ref{fig:mwlsed}.
Both scenarios can reproduce relatively well the IC peak. 
The gamma-ray data of MAGIC and \fermi\ are well explained as the high-energy part of the EC component, with the exception of the two highest energy \fermi\ points which are $>1$\,GeV and hence are probably underestimated by the data selection procedure (see Section~\ref{sec:fermi}).
\textit{Swift}-XRT and historical \textit{Swift}-BAT data follow the rising part of the EC component (with a small contribution of the SSC process in the soft X-ray range for the ``close'' scenario).
The UV data form a bump that can be well explained by the direct accretion disk radiation included in the model. 
In the IR range, the model curve underestimates the data points, especially in the case of the ``far'' model. 
Among the quiescent data selected, the IR data show the highest variability.
The higher IR variability might come from a separate region, not associated with the GeV gamma-ray emission region.

In such a case, the IR emission associated with the low gamma-ray state would likely be at the level closer to the low edge of the observed spread in IR fluxes (reflected in the quoted uncertainty bar in Fig.~\ref{fig:mwlsed}).

The ``far'' model can reasonably reproduce the radio observations, while the ``close'' model underestimates the data due to strong synchrotron-self-absorption effects given by the compactness of the emission region. 
This is not surprising since the radio core observed at 15\,GHz is estimated to be located at the distance of 17.7\,pc from the base of the jet \citep{pu12}. 
Using the typical scaling of the core distance being inversely proportional to the frequency, we obtain that for the highest radio point at 229\,GHz its corresponding radio core should be located at $\sim1$\,pc. 
Therefore, most of the radio emission should be produced at or beyond the region considered in the ``far'' scenario. 
However, the magnetic field considered in the ``far'' scenario, $B=0.05$\,G, is an order of magnitude smaller that the magnetic field estimated from the radio observations at $r=1$\,pc of $0.73$\,G \citep{pu12}.
Larger values would result in a much smaller Compton dominance than observed in the broadband SED.  

It is curious that an optical/GeV high state, a days-long flare and the low state can all be roughly described (except of the IR data) in the framework of the same External Compton scenario without a major change of the model parameters.
This suggests a common origin of the gamma-ray emission of \src\ in different activity states,  with the observed differences caused by changes in the content of the plasma flowing through the emission region\footnote{Note that a fast flare observed in 2016 from \src\ \citep{za17} might have nevertheless a different origin.}.
We note, however, that the model used here is rather simple.
It is natural to assume that the low-state, broad-band emission is an integral of the emission in a range of distances from the base, with the varying conditions (such as $B$ field) along the jet, rather than originating in a single homogenous region (see e.g. \citealp{pc13}). 
  
\subsection{\kom{Limits on the absorption of sub-TeV photons in BLR}}
\begin{figure}[t]
\centering
\includegraphics[width=0.49\textwidth]{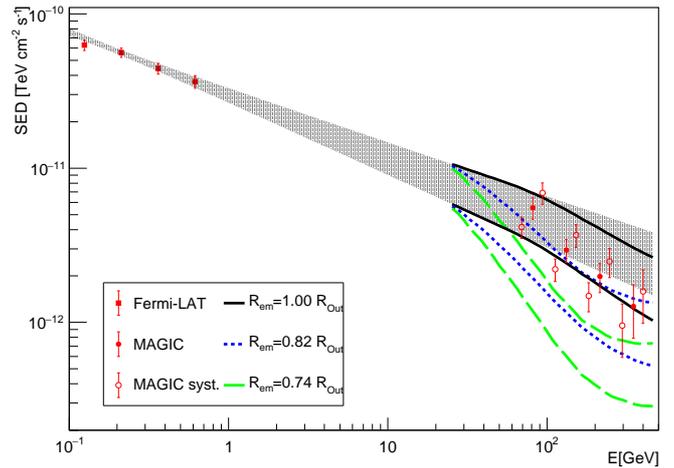}
\caption{
  \kom{Gamma-ray spectrum of \src\ during low state measured by \fermi\ (squares) and MAGIC (filled circles).    
    68\% confidence band of the extrapolation of the \fermi\ spectrum to sub-TeV energies is shown with gray shaded region.
    The extrapolation in the MAGIC energy range assuming absorption in BLR following \citep{be16} for the emission region located at the distance of $1$, $0.82$ and $0.74$ of the outer radius of the broad line region is shown with black solid, blue dotted and green dashed lines respectively.
    Empty circles show the effect of the systematic uncertainties on the MAGIC spectrum. 
    }
}\label{fig:gammatau}
\end{figure}

\kom{In the case the emission region is located inside, or close to, the BLR the gamma-ray spectrum should carry an imprint of the absorption feature on the BLR photons \citep{dp03}. 
Presence or lack of such an absorption can be therefore used to constrain the location of the emission region. 
We use the emission-model-independent approach of \cite{ce17} to put such constraints of the location of the low-state emission region of \src .
We first make a power-law fit to the \fermi\ spectrum of \src\ in the energy range of 0.1--1\,GeV, which is unbiased by the data selection.
Next, we extrapolate the fit to the energy range observed by the MAGIC telescopes and apply an absorption by a factor of $\exp(-\tau)$, where $\tau$ is the so-called optical depth (see Fig.~\ref{fig:gammatau}).
}

\kom{
  We compare those extrapolations with the reconstructed MAGIC spectrum taking into account the statistical uncertainties as well as the systematic uncertainty both in the energy scale and in the flux normalization. 
  The systematic uncertainties of \fermi\ are negligible in those calculations. 
  Due to source intrinsic effects (e.g. intrinsic break or cut-off in the accelerated electron spectrum, Klein-Nishina effect) the sub-TeV spectrum might be below the GeV extrapolation.
  Hence no lower limit on the absorption can be derived in a model-independent way.
  However it is natural to assume that there is no upturning in the photon spectrum, therefore we can place an upper limit on the maximum absorption in the BLR. 
  To estimate such a limit we perform a toy Monte Carlo study in which we vary 10000 times the extrapolated \fermi\ flux and the measured MAGIC flux (corrected for the EBL absorption) according to their uncertainties.
  Next, at each investigated energy we calculate a histogram of $\tau(E)=\ln(F_{\mathrm{ext}}(E)/F_{\mathrm{obs}}(E))$, where
  $F_{\mathrm{ext}}(E)$ and $F_{\mathrm{obs}}(E)$ are the randomized extrapolated and randomized measured flux respectively. 
  The limit on $\tau$ is obtained as a 95\% quantile of such a distribution. 
  We include the systematic uncertainties of MAGIC by shifting its energy scale and normalization according to their systematic limits (see empty circles in Fig.~\ref{fig:gammatau}) and taking the least constraining one.
  We obtain that $\tau(\mathrm{110\,GeV})$<1.4, $\tau(\mathrm{180\,GeV})$<1.7, $\tau(\mathrm{290\,GeV})$<2.3. 
}

  \kom{
  Applying a model of absorption by the BLR those limits on the optical depth can be converted to lower limits on the location of the emission region, $r$.
  We test the above procedure using two BLR models for \src .
  First, we use the optical depth calculations of \cite{be16} assuming that 10\% of the accretion disk radiation is reprocessed in the BLR. 
  Note that \cite{be16} assumes that a homogeneous BLR in \src\ spans between $6.9\times 10^{17}$\,cm and $8.4\times 10^{17}$\,cm, and reflects 10\% of the disk luminosity $L_D=10^{46}\,\mathrm{erg\,s^{-1}}$. 
  We obtain that the above limits result in $r>6.3 \times 10^{17}$\,cm (i.e. above $0.74$ of the outer radius of the BLR).
  As an additional check we calculate the optical depths using a code adapted from \cite{sb08} with the line intensities and BLR geometry of \cite{lb06}. 
  We use the same radius and luminosity of the BLR as in Section~\ref{sec:model}.
  We apply however the same ratio of the outer to inner radius of BLR as in \cite{be16}, resulting in the BLR spanning from $2.34\times 10^{17}$\,cm to $2.86\times 10^{17}$\,cm.
  For such a BLR model, the obtained above limits on the optical depth force us to place the emission region farther than $3.2\times 10^{17}$\,cm (i.e. beyond $1.1$ of the outer radius of the BLR.
}

\kom{
  It should be noted that the method has a number of simplifications and underlying assumptions. 
  The emission region is assumed to be relatively small compared to its distance from the black hole. 
  This is not necessarily true, in particular for the low state emission that can be generated in a more extended region (the broad band emission modelling presented in the previous section further supports such a hypothesis).
  Second, if the gamma-ray emission is not produced by a single process, the spectrum can have a complicated (including convex) shape.
  Note that for another FSRQ, B0218+357 the gamma-ray emission was explained as combination of SSC and EC process \citep{ah16}.
  Third, the optical depth is dependent on the assumed geometry of the BLR.
  For example, the size of the BLR derived in \citep{be16} is a factor of $\sim 3$ larger than the one of \citet{gt09}. 
  In addition the difference between the spherical and disk like geometry can easily change the optical depths by a factor of a few \citep{ap17}.
  Finally, the radial stratification of the BLR and the total fraction of the light reflected in the BLR introduce further uncertainties. 
}

\section{Conclusions}\label{sec:conc}
We have performed the analysis of MAGIC data searching for a possible low state of VHE gamma-ray emission from \src . 
Selecting the data taken during periods when the \fermi\ flux above 1\,GeV was below $3\times10^{-8}\mathrm{cm^{-2}s^{-1}}$ we have collected 75\,hrs of MAGIC data on 76 individual nights, resulting in a significant detection of the low state of VHE gamma-ray emission.  
The measured flux is $\sim0.6$ of the flux of the source measured during high optical and GeV state \citep{al14} in the beginning of 2012 and $\sim0.75$ of the lowest previously known flux from this source (average over all the H.E.S.S. observations, \citealp{za16}).
Nevertheless, the spectral shape is consistent with the previous measurements, despite a factor of 80 difference to the flux during the strongest flare observed so far from this source. 
This makes \src\ the first FSRQ to be detected in a persistent low state with no hints of yearly variations in the observed flux. 
Future observations with the Cherenkov Telescope Array should be able to probe if the low-state flux is also stable on shorter time scales \citep{ach13,ach17}.

\kom{Previous VHE gamma-ray observations of FSRQs in flaring states suggested that the emission region during such states should be located beyond the BLR and that the emission is mostly compatible with an EC scenario.}
The low-state broadband emission \kom{of \src} from IR to VHE gamma-rays can be explained in the framework of an EC scenario, similarly to the previous VHE gamma-ray detections of the source. 
The presence of the sub-TeV gamma-ray emission \kom{also} suggests that the emission region is located beyond the BLR, where the dominating seed radiation field for the EC process is the the dust torus.
\kom{Comparison of the extrapolated \fermi\ spectrum and the MAGIC measured spectrum using two distinct BLR models allows us to put a limit on the location of the emission region beyond the $0.74$ of the outer radius of BLR. 
The emission scenario placing the dissipation region beyond the BLR is in} line with the recent studies of \cite{co18} showing that most of the \fermi\ detected blazars (including \src ) have GeV emission consistent with lack of BLR absorption. 

\begin{acknowledgements}
%
%
We would like to thank the Instituto de Astrof\'{\i}sica de Canarias for the excellent working conditions at the Observatorio del Roque de los Muchachos in La Palma. The financial support of the German BMBF and MPG, the Italian INFN and INAF, the Swiss National Fund SNF, the ERDF under the Spanish MINECO (FPA2015-69818-P, FPA2012-36668, FPA2015-68378-P, FPA2015-69210-C6-2-R, FPA2015-69210-C6-4-R, FPA2015-69210-C6-6-R, AYA2015-71042-P, AYA2016-76012-C3-1-P, ESP2015-71662-C2-2-P, CSD2009-00064), and the Japanese JSPS and MEXT is gratefully acknowledged. This work was also supported by the Spanish Centro de Excelencia ``Severo Ochoa'' SEV-2012-0234 and SEV-2015-0548, and Unidad de Excelencia ``Mar\'{\i}a de Maeztu'' MDM-2014-0369, by the Croatian Science Foundation (HrZZ) Project IP-2016-06-9782 and the University of Rijeka Project 13.12.1.3.02, by the DFG Collaborative Research Centers SFB823/C4 and SFB876/C3, the Polish National Research Centre grant UMO-2016/22/M/ST9/00382 and by the Brazilian MCTIC, CNPq and FAPERJ.

The \textit{Fermi} LAT Collaboration acknowledges generous ongoing support
from a number of agencies and institutes that have supported both the
development and the operation of the LAT as well as scientific data analysis.
These include the National Aeronautics and Space Administration and the
Department of Energy in the United States, the Commissariat \`a l'Energie Atomique
and the Centre National de la Recherche Scientifique / Institut National de Physique
Nucl\'eaire et de Physique des Particules in France, the Agenzia Spaziale Italiana
and the Istituto Nazionale di Fisica Nucleare in Italy, the Ministry of Education,
Culture, Sports, Science and Technology (MEXT), High Energy Accelerator Research
Organization (KEK) and Japan Aerospace Exploration Agency (JAXA) in Japan, and
the K.~A.~Wallenberg Foundation, the Swedish Research Council and the
Swedish National Space Board in Sweden.
 
Additional support for science analysis during the operations phase is gratefully 
acknowledged from the Istituto Nazionale di Astrofisica in Italy and the Centre 
National d'\'Etudes Spatiales in France. This work performed in part under DOE 
Contract DE-AC02-76SF00515.

This paper has made use of up-to-date SMARTS optical/near-infrared light curves that are available at \url{www.astro.yale.edu/smarts/glast/home.php}.

IA acknowledges support by a Ramón y Cajal grant of the Ministerio de Economía, Industria, y Competitividad (MINECO) of Spain.
Acquisition and reduction of the POLAMI and MAPCAT data was supported in part by MINECO through grants AYA2010-14844, AYA2013-40825-P, and AYA2016-80889-P, and by the Regional Government of Andalucía through grant P09-FQM-4784.
The POLAMI observations were carried out at the IRAM 30m Telescope.
IRAM is supported by INSU/CNRS (France), MPG (Germany) and IGN (Spain).
The MAPCAT observations were carried out at the German-Spanish Calar Alto Observatory, which is jointly operated by the Max-Plank-Institut f\"ur Astronomie and the Instituto de Astrofísica de Andalucía-CSIC

This research has made use of data from the OVRO 40-m monitoring program \citep{ri11} which is supported in part by NASA grants NNX08AW31G, NNX11A043G, and NNX14AQ89G and NSF grants AST-0808050 and AST-1109911.

This publication makes use of data obtained at the Mets\"ahovi Radio Observatory, operated by Aalto University, Finland.
The authors would like to thank a number of people who provided comments to the manuscript: R.~Angioni, N.~MacDonald, M.~Giroletti, R.~Caputo, D.~Thompson and the anonymous journal reviewer. 
\kom{We would also like to thank M.~B\"ottcher for providing us the numerical values of the optical depths calculated in \cite{be16}.}
\end{acknowledgements}

%
%

\appendix
\section{Data selection tests}\label{app:datasel}
As discussed in Section~\ref{sec:fermi} the energy threshold of 1\,GeV used for the selection was motivated by its proximity to the VHE energy range.
However, the daily estimation of the flux above 1\,GeV have large uncertainty which can in principle affect the data selection.
To test this effect we have applied instead a cut to select nights with flux above 100\,MeV measured by \fermi\ to be below $10^{-6} \mathrm{cm^{-2}s^{-1}}$.
Such a cut results in the same number of the MAGIC observations nights (76) selected for the low state analysis.
The number of individual nights that would change the classification to the low state or to the high state with such a cut is 9 (corresponding to 12\% of the low-state sample) each. 
We have tested the validity of the used here low-state data selection procedure by applying a cut at the flux above 100\,MeV instead of 1\,GeV.
We therefore conclude, that the value of the \fermi\ analysis threshold do not have a large impact on the selection of nights used for this analysis. 
We have tested the effect of leaving the spectral index free in the light curve analysis, and found that this does not strongly affect the fraction of the data which results to be classified as low GeV state, following the definition described above.
Fixing the spectral index to the average value of $2.36$ (see the 3FGL catalog, \citealp{Acero15}) would change the number of nights assigned to the low state and high state by $\lesssim1\%$ each.

\end{document}